\documentclass[lettersize,journal]{IEEEtran}

\usepackage{hyperref}
\usepackage{amsmath,amsfonts}
\usepackage{physics,amsmath}
\usepackage{amssymb}
\usepackage{algorithm}
\usepackage{algpseudocode}
\usepackage{array}
\usepackage{bm}
\usepackage{subcaption}

\usepackage{mathtools}
\usepackage{xcolor}
\usepackage{tikz}
\usetikzlibrary{arrows.meta, positioning, shapes.geometric, shadows}
\usepackage{subcaption}
\usepackage{textcomp}
\usepackage{stfloats}
\usepackage{lipsum}  
\usepackage{url}
\usepackage{verbatim}
\usepackage{graphicx}
\usepackage{cite}
\usepackage{pdfpages}
\usepackage{tikz}
\usetikzlibrary{calc}
\usetikzlibrary{3d}
\usetikzlibrary{arrows.meta} 

\DeclareCaptionLabelFormat{simple}{#2}
\begin{document}

\title{
Chemo-Hydrodynamic Transceivers for the Internet of Bio-Nano Things, Modeling the Joint Propulsion Transmission trade-off}
\author{Shaojie Zhang,~\IEEEmembership{Student             Member,~IEEE}
        and Ozgur B. Akan,~\IEEEmembership{Fellow,~IEEE}
        \thanks{Shaojie Zhang is with the Internet of Everything Group, Electrical
        Engineering Division, Department of Engineering, University of Cambridge,
        CB3 0FA Cambridge, U.K. (e-mail: sz466@cam.ac.uk). }
        \thanks{Ozgur B. Akan is with the Internet of Everything Group, Electrical
        Engineering Division, Department of Engineering, University of Cambridge,
        CB3 0FA Cambridge, U.K., and also with the Center for neXt-Generation
        Communications (CXC), Department of Electrical and Electronics Engineering, Ko\text{\c{c}} University, 34450 Istanbul, Turkey (e-mail: oba21@cam.ac.uk;
        akan@ku.edu.tr)}
        \thanks{This work was supported in part by the AXA Research Fund (AXA Chair
for Internet of Everything at Ko\text{\c{c}} University).}
}


\maketitle

\begin{abstract}
The Internet of Bio-Nano Things (IoBNT) requires mobile nanomachines that navigate complex fluids while exchanging molecular signals under external supervision. We introduce the chemo-hydrodynamic transceiver, a unified model for catalytic Janus particles in which an external optical control simultaneously drives molecular emission and active self-propulsion. Unlike common abstractions that decouple mobility and communication, we derive a stochastic channel model that captures their physicochemical coupling and shows that actuation-induced distance jitter can dominate the received-signal variance, yielding a fundamental trade-off: stronger actuation increases emission but can sharply reduce reliability through motion-induced fading. Numerical results reveal a unimodal reliability profile with a critical actuation level beyond which the signal-to-noise ratio collapses, and an optimal control level that scales approximately linearly with link distance. Compared with Brownian-mobility baselines, the model exposes a pronounced estimation gap: neglecting active motility noise can underestimate the bit error probability by orders of magnitude. These findings provide physical-layer guidelines for mobility-aware IoBNT protocol design and closed-loop control of nanorobotic swarms.
\end{abstract}

\section{Introduction}
\begin{figure}[t]
    \centering
    \includegraphics[width = 0.55\textwidth]{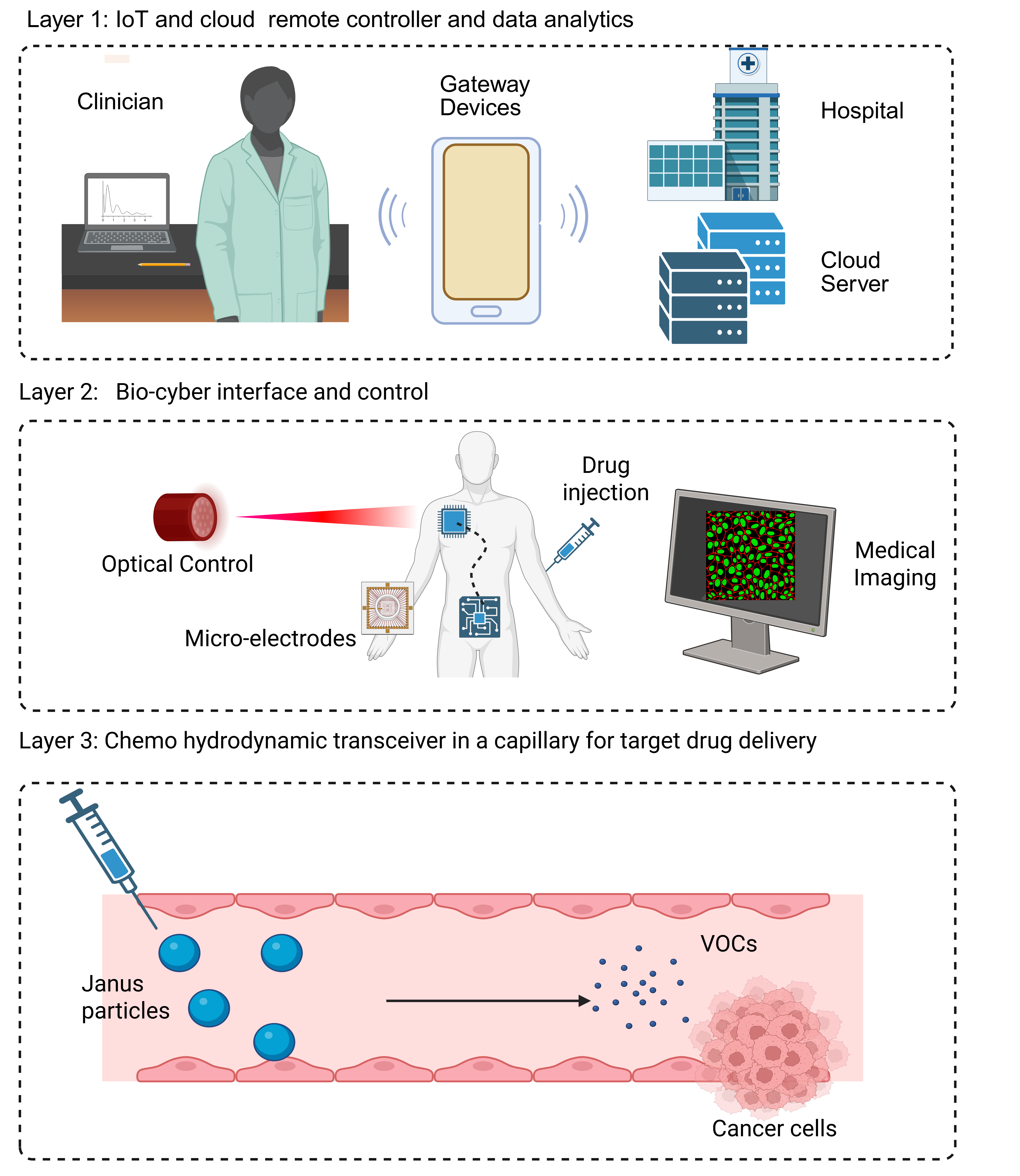}
    \caption{Closed-loop IoBNT architecture in which the external command $I_{\text{ext}}(t)$ simultaneously governs propulsion $U(t)$ and molecular transmission, making channel statistics dependent on transceiver motility.}
\label{fig:system_scenario}
\end{figure}

The Internet of Bio-Nano Things (IoBNT) extends IoT concepts to fluidic and physiological environments, enabling networked sensing and actuation in settings such as the vasculature and aquatic systems \cite{Akyildiz2015internet,Zafar2021IoBNT}. While recent IoBNT research has predominantly focused on static biosensing and passive diffusion based communication links \cite{Nakano2012molecular}, the realization of mission-critical applications in nanomedicine—including targeted drug delivery and high resolution \textit{in situ} biopsy—necessitates the deployment of mobile agents capable of autonomous navigation, sensing, and actuation under macro-scale supervision \cite{ChudeOkonkwo2017TDD, AlZubi2022intrabody}. Within these emerging bio-cyber interfaces, mobile nanomachines function as autonomous edge nodes within a closed loop cyber-physical system, wherein external control commands must be reliably translated into both hydrodynamic motion and molecular information exchange \cite{Akyildiz2015internet,Zafar2021IoBNT}.

A key system-design challenge is developing physical-layer models that map macro-scale control inputs (e.g., optical intensity) to link statistics that govern reliability, latency, and control-loop stability \cite{Nakano2012molecular,AlZubi2022intrabody}. However, conventional molecular communication abstractions were primarily formulated for static transmitters or passive carriers; consequently, their direct application to mobility aware IoBNT protocol design may yield erroneous design conclusions. We identify two prevalent modeling limitations in the existing literature: the reservoir abstraction and the mobility decoupling assumption \cite{Nakano2012Opportunities,Jamali2019Tutorial}. The reservoir abstraction treats molecular emission as an independent design parameter, decoupled from the device physics and motility state \cite{Pierobon2010EndToEnd,Jamali2019Tutorial,Meng2014Receiver}. Conversely, the mobility decoupling assumption treats trajectories as fixed a priori, typically via Brownian motion, and assumes mobility statistics are independent of the transmission strategy \cite{Ahmadzadeh2018Stochastic,Chang2018Adaptive,Varshney2023RescaledBM}. For closed loop IoBNT operations, these assumptions can distort link budgeting, rate selection, and actuation policies, potentially leading to unstable control decisions where increased actuation, intended to improve communication, paradoxically amplifies channel uncertainty.

This article addresses chemo-hydrodynamic transceivers, a class of catalytic nanomotors wherein actuation and communication are intrinsically coupled. Specifically, we consider catalytic Janus particles that propel via self diffusiophoresis, a mechanism where an external input modulates a surface reaction that concurrently generates a propulsive slip velocity and a flux of signaling molecules \cite{Moran2017PhoreticReview,Yang2016ConfinedJCM,Mozaffari2016BoundaryDiffusiophoresis}. Recent realizations of light responsive swimmers support a direct transduction interface in which a single control signal simultaneously shapes the propulsion speed $U$ and the emission process, thereby altering the effective diffusion $D_{eff}$ of the transmitter in a signal dependent manner \cite{Feldmann2019LightJanus,Yu2019Phototaxis,Zhou2019PulsatingJanus}. Consequently, the channel noise statistics exhibit a dependency on the control input, such that aggressive actuation strategies may induce self inflicted stochasticity that dominates link reliability. This coupling necessitates a unified modeling framework suitable for IoT-oriented design inquiries, including mobility aware link adaptation, symbol duration optimization, and the definition of safe actuation envelopes for reliable cyber-physical operation. Throughout, we focus on a single-link setting under quasi-steady concentration response and negligible inter-symbol interference (ISI) to isolate this fundamental trade-off.

The specific contributions of this paper are summarized as follows:
\begin{enumerate}
    \item \textbf{Control-Dependent Mobility, Actuation Trade-off, and Estimation Gap:} We introduce a system-level model for mobile IoBNT nodes where a single external input governs both propulsion and molecular emission, rendering the mobility statistics control dependent. This yields a closed-form actuation envelope that quantifies how increasing control effort amplifies motion-induced fading and degrades link reliability. We also quantify an estimation gap, showing that standard mobile diffusion models can underestimate the BEP by orders of magnitude in high-mobility regimes.
    \item \textbf{End-to-End Stochastic Channel Law with Signal-Dependent Variance:} By embedding microscopic phoretic relations \cite{DeGraaf2015diffusiophoretic} into mesoscopic active Brownian dynamics \cite{Howse2007Selfmotile}, we derive a tractable observation model. The signal mean scales linearly with the control level, while the variance incorporates an active component that scales quartically with the control input.
    \item \textbf{Reliability Analysis and Closed-Form Design Rules:} For binary signaling, we derive closed-form expressions for the Signal-to-Noise Ratio (SNR) and Bit Error Probability (BEP) that establish a non-monotonic reliability behavior. The analysis proves the existence of an optimal actuation level and yields an explicit design law for the optimal control intensity with distance scaling, along with constraints on symbol duration $T$ due to accumulated motion-induced jitter.
\end{enumerate}

The remainder of this paper is organized as follows. Section II establishes the deterministic chemo-hydrodynamic actuation model. Section III derives the stochastic propagation dynamics. Section IV and Section V present the receiver observation model and the error probability analysis, respectively. Section VI provides numerical validation via stochastic particle based simulations and discusses the implications for system design. Finally, Section VII concludes the paper.

\section{Chemo-Hydrodynamic Transceiver Model}
\label{sec:chemo_hydro_transceiver}
\begin{figure*}[t]
\centering
\resizebox{\textwidth}{!}{%
\begin{tikzpicture}[>=latex, font=\large, node distance=1.2cm]

    \tikzset{
        block/.style={
            draw,
            rectangle,
            rounded corners,
            minimum height=1.3cm,
            minimum width=3.0cm,
            align=center,
            fill=white,
            thick
        },
        sum/.style={draw, circle, inner sep=0pt, minimum size=6mm},
        smallnote/.style={font=\large} 
    }

    \node[
        draw,
        rounded corners,
        fill=yellow!20,
        thick,
        minimum height=1.0cm,
        minimum width=2.6cm,
        align=center
    ] (input) {Control input\\$I_{\text{ext}}(t)$};

    \node[block, right=2.2cm of input, fill=blue!5] (micro) {
        \textbf{Chemo-hydrodynamics}\\[1pt]
        Surface reaction\\
        Slip velocity
    };
    \node[smallnote, below=0.05cm of micro] {Sec.~\ref{sec:chemo_hydro_transceiver}};

    \node[block, right=2.2cm of micro, fill=orange!5] (meso) {
        \textbf{Active mobility}\\[1pt]
        Langevin dynamics\\
        Effective diffusion
    };
    \node[smallnote, below=0.05cm of meso] {Sec.~\ref{sec:stochastic_propagation}};

    \node[block, right=2.2cm of meso, fill=green!5] (macro) {
        \textbf{Channel statistics}\\[1pt]
        Linear observation\\
        Signal dependent noise
    };
    \node[smallnote, below=0.05cm of macro] {Sec.~\ref{sec:receiver_model}};

    \node[sum, right=1.6cm of macro] (adder) {+};
    \node[right=1.4cm of adder] (output) {};
    \node[above=0.9cm of adder] (measnoise) {$Z_m \sim \mathcal{N}(0,\sigma_m^2)$};

    \draw[->, thick]
        (input) -- node[midway, below] {$I_{\text{ext}}(t)$} (micro);

    \draw[->, thick]
        (micro) -- node[midway, below] {$U(t)$} (meso);

    \draw[->, thick]
        (meso) -- node[midway, below] {$D_{\text{eff}}(t)$} (macro);

    \draw[->, thick] (macro) -- (adder);
    \draw[->]        (measnoise) -- (adder);

    \draw[->, thick]
        (adder) -- node[midway, below] {$Y(t)$} (output);

    \coordinate (mid)    at ($(micro.south)!0.5!(macro.south)$);
    \coordinate (midlow) at ($(mid)-(0,1.5)$);

    \draw[->, thick, blue!60!black]
        (micro.south) |- (midlow)
        -- (midlow -| macro.south)
        node[midway, below] {Emission $q(t)$}
        -- (macro.south);

    \node[above=1.2cm of meso, align=center] (thermal) {Thermal fluctuations\\$\xi(t) \sim \mathcal{N}(0,1)$};
    \draw[->, dashed] (thermal) -- (meso);

    \node[above=0.7cm of micro, text=blue!60!black,   font=\bfseries] {Micro scale};
    \node[above=0.7cm of meso,  text=orange!60!black, font=\bfseries] {Meso scale};
    \node[above=0.7cm of macro, text=green!60!black,  font=\bfseries] {Macro scale};

\end{tikzpicture}%
}
\caption{Multi-scale modeling roadmap: $I_{\text{ext}}(t)$ drives micro-scale surface reaction and slip, setting propulsion $U(t)$ and emission $q(t)$. Propulsion enters meso-scale stochastic dynamics, yielding a time-varying effective diffusion $D_{\text{eff}}(t)$. At the macro scale, $q(t)$ and $D_{\text{eff}}(t)$ determine channel statistics and the received signal $Y(t)$ with signal-dependent variance.}
\label{fig:modeling_roadmap}
\end{figure*}
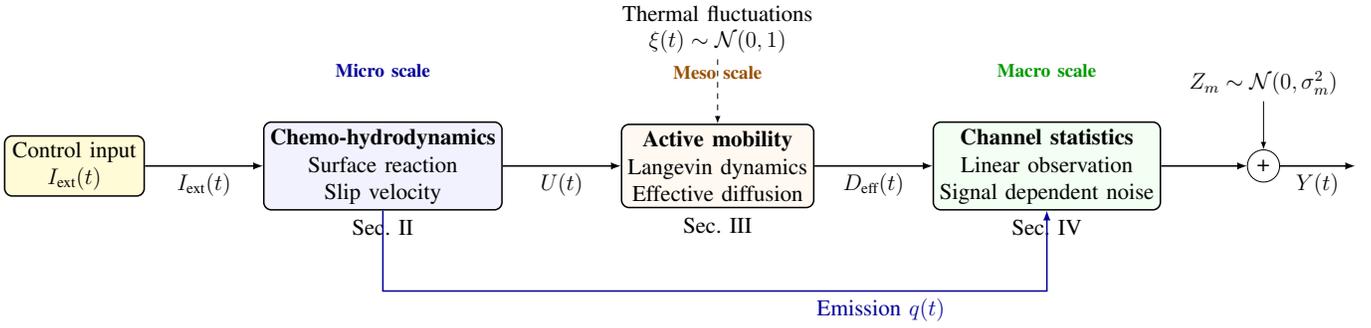

In this section, We derive a compact mapping from the external control signal $I_{\text {ext }}(t)$ to propulsion $U(t)$ and emission rate $q(t)$, which forms the input stage of the end-to-end channel model in Fig.~\ref{fig:modeling_roadmap}.

\begin{figure}[t]
    \centering
    \includegraphics[width = 0.45\textwidth]{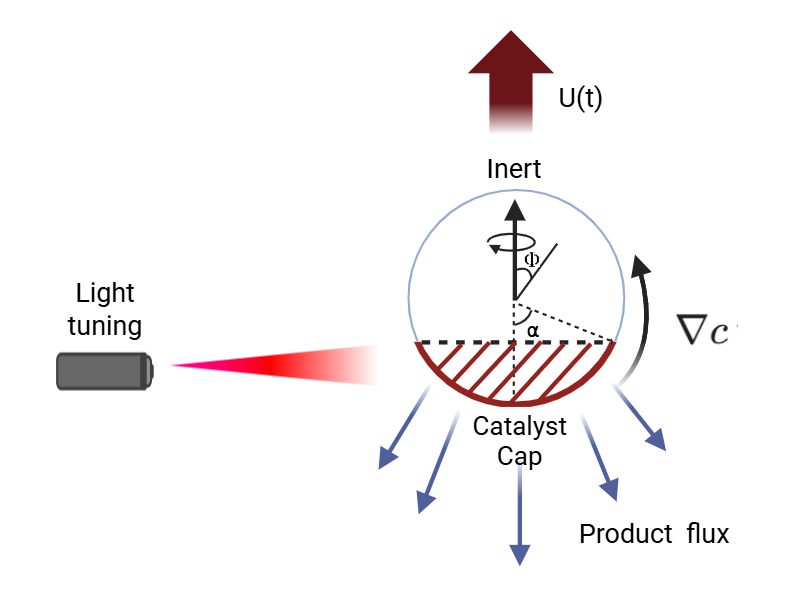}
    \caption{Cross-sectional schematic of the chemo-hydrodynamic transceiver. The external input $I_{\text{ext}}(t)$ modulates the reaction flux $k(\theta)$ on the catalytic hemisphere ($\theta<\alpha$), creating a concentration gradient $\nabla c$ that drives phoretic slip $\mathbf{v}_s$ and net propulsion $\mathbf{U}(t)$. The same surface reaction also sets the emission rate $q(t)$ of information molecules, coupling mobility and communication.}
\label{fig:chemo_hydro_micro}
\end{figure}

\subsection{Geometry and Physical Assumptions}
\label{subsec:geometry}

We consider a rigid spherical particle of radius $a$ immersed in a Newtonian fluid characterized by dynamic viscosity $\eta$.  The particle surface comprises an inert hemisphere and a catalytic cap defined by the polar angle $0 \leq \theta \leq \alpha$ where $\theta=0$ aligns with the symmetry axis $\hat{\mathbf{e}}_z$.

The fluid contains a fuel solute designated as species $A$ with diffusion coefficient $D$. We model the chemical mechanism as an irreversible surface reaction on the catalytic cap,
\begin{equation}
A \longrightarrow B,
\end{equation}
where the consumption of fuel $A$ drives self propulsion and the generated product $B$ is used as the information carrier.

The external control input $I_{\text{ext}}(t)$ modeled here as a dimensionless normalized intensity modulates the local surface consumption flux of the fuel species $A$ on the cap,
\begin{equation}
    k(\theta,t) = \kappa_{\text{base}}\, I_{\text{ext}}(t)\, \Theta(\alpha - \theta),
    \label{eq:surface_flux_control}
\end{equation}
where $k(\theta,t)$ has units of (molecules or moles)$\,\text{m}^{-2}\,\text{s}^{-1}$ and $\kappa_{\text{base}}$ is the baseline flux amplitude (same units) at unit intensity. Here $\Theta(\cdot)$ represents the Heaviside step function. 
We assume a linear response regime (negligible heating and fuel depletion). Saturation at high intensity can be captured by replacing $I_{\text {ext }}(t)$ with a monotone saturation function. Our analysis uses far-field linearization $|x| / d \ll 1$, low Péclet number, and quasi-steady diffusion.

\subsection{Solute Diffusion and Diffusiophoretic Slip}
\label{subsec:solute_diffusion}

The reaction defined in \eqref{eq:surface_flux_control} creates a local depletion of fuel $A$ generating a concentration gradient that drives the particle via self diffusiophoresis. We adopt the standard quasi-steady approximation $\tau_D \ll T$ solving the Laplace equation $\nabla^2 c_A = 0$ for the fuel concentration $c_A$ subject to the flux boundary condition $D \partial_r c_A |_{r=a} = k(\theta,t)$ \cite{DeGraaf2015diffusiophoretic}.

The interaction between the fuel gradient and the particle surface induces a diffusiophoretic slip velocity $\mathbf{v}_s(\theta,t) = - b_{\mathrm{DP}} \nabla_{\parallel} c_A\vert_{r=a}
$ where $b_{\mathrm{DP}}$ denotes the diffusiophoretic mobility parameter \cite{Anderson1989Colloid,Moran2017PhoreticReview}. While gradients of the product species $B$ may also contribute in general, throughout we assume that the slip is dominated by the fuel field $c_A$ and that the contribution of $c_B$ is negligible (e.g., due to a much smaller phoretic mobility with respect to $B$ or weak $B$--surface interactions). This assumption is consistent with simultaneously using $B$ as the information molecule: the receiver observes the far-field release of $B$ (set by reaction stoichiometry), whereas the propulsion is set by the near-field gradient of $A$.

For a force free sphere only the first Legendre mode of this slip profile contributes to net translation \cite{Moran2017PhoreticReview}. As detailed in Appendix~\ref{app:chemo_derivation} the effective slip dipole coefficient $B_1(t)$ corresponding to this mode is
\begin{equation}
    B_1(t) = - \frac{3b_{\mathrm{DP}}}{8D}\, \sin^2\alpha\, \kappa_{\text{base}}\, I_{\text{ext}}(t).
    \label{eq:B1_main}
\end{equation}
This coefficient $B_1(t)$ encapsulates the polarity of the self generated concentration field driven by the control input.

\subsection{Propulsion Velocity and Control Gain}
\label{subsec:propulsion_gain}

Using the reciprocal theorem for Stokes flow \cite{stone1996propulsion} the net swimming velocity is determined as $U(t) = \frac{2}{3} B_1(t)$. Substitution of \eqref{eq:B1_main} yields a linear control law
\begin{equation}
    U(t) = K_{\text{control}}\, I_{\text{ext}}(t),
    \label{eq:U_linear_main}
\end{equation}
where the aggregate propulsion gain $K_{\text{control}}$ is defined as
\begin{equation}
    K_{\text{control}} = - \frac{b\, \kappa_{\text{base}}}{4D} \sin^2\alpha.
    \label{eq:Kcontrol_main}
\end{equation}
The gain $K_{\text{control}}$ aggregates the geometric $\alpha$ chemical $\kappa_{\text{base}}$ and hydrodynamic parameters $b_{\mathrm{DP}}$ and $D$. For repulsive interactions where $b < 0$ the gain $K_{\text{control}}$ is positive indicating propulsion along the symmetry axis.

Simultaneously, the emission rate $q(t)$ of the information molecule species $B$ is stoichiometrically linked to the same surface reaction. Specifically, integrating the fuel consumption flux over the catalytic cap yields the total consumption rate of $A$ (molecules$\,\text{s}^{-1}$), yielding the production rate of $B$,
\begin{equation}
    q(t)=\int_{\partial\Omega} k(\theta,t)\,\mathrm{d}A
    = \kappa_{\text{base}}\, A_{\text{cap}}\, I_{\text{ext}}(t),
\end{equation}
where $A_{\text{cap}} = 2\pi a^2(1-\cos\alpha)$ is the catalytic cap area. Accordingly, we identify the (control-to-emission) gain
\begin{equation}
    \kappa_{\text{em}} \triangleq \kappa_{\text{base}}\, A_{\text{cap}},
\end{equation}
so that $q(t)=\kappa_{\text{em}} I_{\text{ext}}(t)$.

In Section \ref{sec:stochastic_propagation} we will utilize $U(t)$ as the drive parameter for an active Brownian motion model where it induces an input dependent effective diffusion that fundamentally alters the channel statistics.



\section{Stochastic Propagation Model}
\label{sec:stochastic_propagation}

In this section we establish the stochastic mobility block of the transceiver model. We extend the deterministic propulsion framework from Section~\ref{sec:chemo_hydro_transceiver} to account for the intrinsic randomness of micro scale motion. The objective is to derive the statistical distribution of the particle's axial position $x(t)$ as a function of the control input $I_{\text{ext}}(t)$ which serves as the input to the receiver observation model.

\subsection{Planar Active Brownian Dynamics}
\label{subsec:planar_dynamics}

We assume the particle motion is constrained to a two dimensional plane consistent with typical experimental configurations for catalytic swimmers \cite{Howse2007Selfmotile,TenHagen2011Brownian}. The instantaneous state is characterized by the position vector $\mathbf{r}(t) = [x(t), y(t)]^\top$ and the orientation angle $\varphi(t)$. The stochastic evolution of these variables is governed by the coupled overdamped Langevin equations \cite{TenHagen2011Brownian}
\begin{align}
    \frac{\mathrm{d}\mathbf{r}}{\mathrm{d}t} &= U(t) \begin{pmatrix} \cos\varphi \\ \sin\varphi \end{pmatrix} + \sqrt{2 D_t}\, \boldsymbol{\xi}_t(t), \label{eq:langevin_pos}\\
    \frac{\mathrm{d}\varphi}{\mathrm{d}t} &= \sqrt{2 D_r}\, \xi_{\varphi}(t), \label{eq:langevin_angle}
\end{align}
where $U(t) = K_{\text{control}} I_{\text{ext}}(t)$ is the controlled propulsion speed. The terms $\boldsymbol{\xi}_t$ and $\xi_{\varphi}$ represent independent Gaussian white noise processes.  The translational diffusion coefficient $D_t$ and rotational diffusion coefficient $D_r$ follow the Stokes Einstein relations
\begin{equation}
    D_t = \frac{k_B T_{\mathrm{env}}}{6\pi \eta a}, \qquad D_r = \frac{k_B T_{\mathrm{env}}}{8\pi \eta a^3}.
    \label{eq:diff_coeffs}
\end{equation}
The orientation dynamics are characterized by the rotational relaxation time $\tau_r = (D_r)^{-1}$ which dictates the timescale of directional persistence \cite{Howse2007Selfmotile}.

\subsection{Effective 1D Propagation Channel}
\label{subsec:effective_channel}

We adopt a one dimensional confined geometry (e.g., a narrow capillary) such that the particle motion relevant to the communication link is well approximated by its axial displacement $x(t)$ along the transmitter--receiver line. Under this confinement we model the transverse coordinates as suppressed by the channel walls and focus on the statistics of $x(t)$.

We assume that the control signal is 
\emph{piecewise constant per symbol}, i.e., $I_{\text{ext}}(t)=I$ for $t\in[0,T]$, and that the orientation is 
\emph{mixed between symbols} (the orientation at the beginning of each symbol is independent and isotropically distributed). This removes inter-symbol orientation memory and allows each symbol interval to be analyzed independently. Equivalently, we assume a sufficient guard interval between consecutive transmissions to allow rotational decorrelation, or we treat successive symbols as independent realizations for the purpose of deriving a reliability envelope.

For a constant propulsion speed $U_0$ over $[0,T]$, the exact mean squared displacement (MSD) of an active Brownian particle along one axis is \cite{Howse2007Selfmotile}
\begin{equation}
    \langle \Delta x(t)^2 \rangle = 2D_t t + \frac{U_0^2}{D_r^2}(e^{-D_r t} + D_r t - 1).
    \label{eq:msd_1d_axis}
\end{equation}
Therefore, at the decision time $t=T$ the axial position is modeled as zero-mean Gaussian with variance
\begin{equation}
    \sigma_x^2(T) = 2D_t T + \frac{U_0^2}{D_r^2}(e^{-D_r T} + D_r T - 1).
    \label{eq:sigmax_exact_decision}
\end{equation}
Importantly, \eqref{eq:sigmax_exact_decision} is valid for all $T$ (including $T\sim\tau_r$) and eliminates the need to invoke a long-time effective diffusivity. In our controlled setting $U_0 = U = K_{\text{control}} I$, hence $\sigma_x^2(T)$ depends explicitly on the control amplitude $I$ through the $U^2$ term.

\section{Receiver Observation Model}
\label{sec:receiver_model}

In this section we formulate the receiver observation model designed to map the stochastic position of the Janus particle to the statistical properties of the received signal. We develop a discrete time channel model wherein the signal characteristics exhibit an explicit dependence on the control input thereby capturing the intrinsic physicochemical coupling between the propulsion mechanism and the molecular emission process.

\begin{figure}[t]
    \centering
    \resizebox{0.95\columnwidth}{!}{%
    \begin{tikzpicture}[>=stealth, font=\small]
        \fill[gray!10] plot[domain=-1.5:1.5, samples=50] (\x, {1.2*exp(-1.5*\x*\x)}) -- cycle;
        \draw[gray, thin] plot[domain=-1.5:1.5, samples=50] (\x, {1.2*exp(-1.5*\x*\x)});
        \draw[->, thick] (-2.0,0) -- (6.5,0) node[below] {$x$};
        \draw[thick] (0, 0.1) -- (0, -0.1) node[below] {$x=0$};
        \node[gray, above, font=\scriptsize] at (0, 1.3) {PDF $\sim \mathcal{N}(0, \sigma_x^2)$};

        \def\xt{1.2} 
        \def\rad{0.25}
        \draw[dashed] (\xt, 0) -- (\xt, 2.0);
        \begin{scope}[shift={(\xt, 0.35)}] 
            \fill[white] (0,0) circle (\rad);
            \draw[black] (0,0) circle (\rad);
            \begin{scope}
                \clip (0,-\rad) rectangle (-\rad, \rad);
                \fill[gray!80] (0,0) circle (\rad);
            \end{scope}
            \draw[black] (0,0) circle (\rad);
            \draw[->, blue, thick] (0,0) -- (0.5, 0);
        \end{scope}
        \node[fill=white, inner sep=1pt] at (\xt, 2.0) {$x(t)$};
        \node[align=center, font=\scriptsize] (TxLabel) at (0.5, -1.2) {\textbf{Chemo-Hydrodynamic}\\ \textbf{Transceiver}};
        \draw[->, gray] (TxLabel.north) -- (\xt, 0.1);

        \def\d{5.5}
        \draw[fill=blue!10, thick] (\d-0.2, 0) rectangle (\d+0.2, 0.7);
        \node[font=\bfseries\scriptsize, align=center] at (\d, 0.35) {Rx};
        \draw[dashed] (\d, 0.7) -- (\d, 2.0);
        \node[above] at (\d, 2.0) {$x=d$};

        \draw[<->, thick] (\xt, 1.7) -- (\d, 1.7);
        \node[fill=white, inner sep=1pt, font=\footnotesize] at ({(\xt+\d)/2}, 1.7) {$R(t)$};
        \draw[<->, gray] (0, 1.0) -- (\d, 1.0);
        \node[fill=white, inner sep=1pt, font=\scriptsize, text=gray] at ({\d/2}, 1.0) {Link Distance $d$};
    \end{tikzpicture}%
    }
    \caption{Schematic representation of the one dimensional channel geometry. The axial position of the transceiver $x(t)$ is modeled as a stochastic variable characterized by a probability density function that is dynamically determined by the control input. The receiver node is located at a fixed coordinate $x=d$.}
    \label{fig:coord_geometry}
\end{figure}
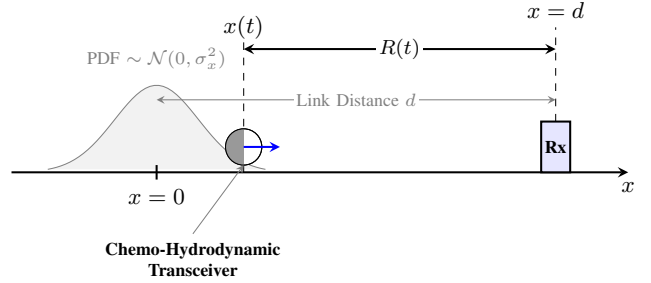

\subsection{Concentration Field and Quasi-Steady Approximation}
\label{subsec:concentration_field}

The catalytic reaction occurring on the particle surface generates solute species $B$ at a rate proportional to the instantaneous control signal $I_{\text{ext}}(t)$
\begin{equation}
    q(t) = \kappa_{\text{em}}\, I_{\text{ext}}(t),
    \label{eq:emission_rate}
\end{equation}
where $\kappa_{\text{em}}$ denotes the emission rate constant. The resulting concentration field $c_B(\mathbf{r},t)$ is governed by the diffusion equation subject to a moving point source located at $x(t)$ \cite{ChudeOkonkwo2017TDD}.

To remain consistent with the instantaneous sampling model in Section~\ref{subsec:linearized_model}, we assume that ISI is negligible. This can be ensured, for example, by enzymatic/chemical degradation of the information molecules or by allocating a sufficient clearing/guard time between consecutive symbols so that the residual concentration from the previous emission is negligible at the next decision instant \cite{Jamali2019Tutorial}.

We adopt a quasi-steady far-field approximation, which is accurate when the diffusion time over the link is not larger than the symbol timescale, i.e., $d^2/D_B \lesssim T$, and when the Péclet number remains low, $Pe = Ud/D_B \ll 1$. Since the propulsion speed $U$ scales with $I_{\text{ext}}$, the Péclet constraint defines the validity range for the analytical model.

We define the instantaneous separation distance as $R(t)$. Assuming that lateral fluctuations $y(t)$ are small relative to the axial distance $d$ or that the receiver integrates flux over a sufficient cross section we approximate $R(t) \approx d - x(t)$. The concentration at the receiver is thus derived as
\begin{equation}
    c_R(t) \approx \frac{G_{\text{ch}}\, I_{\text{ext}}(t)}{d - x(t)},
    \label{eq:quasi_steady_conc}
\end{equation}
where $G_{\text{ch}} = \kappa_{\text{em}} / (4\pi D_B)$. Analytically we assume the probability of the particle approaching the singularity at $x=d$ is negligible under far field operation. In numerical simulations this physical constraint is strictly enforced via a reflective boundary.

\subsection{Linearized Channel Model}
\label{subsec:linearized_model}

We adopt an instantaneous sampling model where the receiver measures the concentration at the decision instant $t=T$ consistent with a standard thresholding receiver. 

The observation $Y$ is modeled as the scaled concentration plus additive noise
\begin{equation}
    Y = \beta_R\, c_R(T) + Z_m = \frac{H_0\, I_{\text{ext}}(T)}{d - X_b} + Z_m,
    \label{eq:Y_nonlinear}
\end{equation}
where $H_0 = \beta_R G_{\text{ch}}$ represents the channel gain and $X_b = x(T)$ denotes the stochastic displacement. The term $Z_m \sim \mathcal{N}(0, \sigma_m^2)$ models the aggregate receiver uncertainty.

Invoking the far field assumption we linearize \eqref{eq:Y_nonlinear} utilizing a first order Taylor expansion. This approximation remains valid when the positional standard deviation is small relative to the link distance quantitatively adhering to the criterion $\sigma_x / d \le 0.1$. In Section VI we confirm that the optimal operating points lie well within this regime and that the particle based simulations utilize the exact nonlinear observation \eqref{eq:Y_nonlinear} to verify that the trends are not artifacts of the linearization
\begin{equation}
    Y \approx \frac{H_0 I_b}{d} \left( 1 + \frac{X_b}{d} \right) + Z_m = \mu_b + \alpha_b X_b + Z_m,
    \label{eq:Y_linearized}
\end{equation}
where $I_b$ is the constant control input for symbol $b$. Here $\mu_b = (H_0/d) I_b$ represents the mean signal and $\alpha_b = (H_0/d^2) I_b$ denotes the motion induced fluctuation coefficient.

\subsection{Signal Dependent Noise Statistics}
\label{subsec:signal_dependent_noise}
Based on Section~\ref{sec:stochastic_propagation}, $X_b=x(T)$ follows $\mathcal{N}(0,\sigma_{x,b}^2)$ with
\begin{equation}
\sigma_{x,b}^2 \triangleq \sigma_x^2(T)\big|_{U_0=U_b}
= 2D_t T + \frac{U_b^2}{D_r^2}\!\left(e^{-D_r T}+D_r T-1\right),
\label{eq:sigma_xb_exact}
\end{equation}
where $U_b = K_{\text{control}} I_b$.
For $T\gg 1/D_r$, this reduces to $\sigma_{x,b}^2 \approx 2(D_t+U_b^2/(2D_r))T$.
\begin{equation}
\mu_b \triangleq \mathbb{E}[Y\mid \mathcal{H}_b] = \frac{H_0}{d}\, I_b .
\label{eq:mu_b}
\end{equation}
Consequently the conditional observation $Y|I_b$ is Gaussian with mean $\mu_b$ and total variance $\sigma_{Y,b}^2$
\begin{equation}
\sigma_{Y,b}^2 = \sigma_m^2 + \alpha_b^2 \sigma_{x,b}^2,
\label{eq:variance_intermediate}
\end{equation}
Substituting \eqref{eq:sigma_xb_exact} into \eqref{eq:variance_intermediate} yields the polynomial form
\begin{equation}
    \sigma_{Y,b}^2 = \underbrace{\sigma_m^2}_{\text{Measurement}}
    + \underbrace{2 \left( \frac{H_0}{d^2} \right)^2 D_t T I_b^2}_{\text{Passive Diffusion}}
    + \underbrace{\left( \frac{H_0}{d^2} \right)^2 K_{\text{control}}^2\, g(T)\, I_b^4}_{\text{Active Propulsion}},
    \label{eq:variance_decomposition}
\end{equation}
where
\begin{equation}
g(T)\triangleq \frac{e^{-D_r T}+D_r T-1}{D_r^2}.
\label{eq:gT_def}
\end{equation}
In the long-time regime $T\gg 1/D_r$, $g(T)\approx T/D_r$, so the active-propulsion term in \eqref{eq:variance_decomposition} reduces to the long-time coefficient used to define $c_3$ in \eqref{eq:SNR_limit}.

Equation \eqref{eq:variance_decomposition} reveals the critical limitation of chemo-hydrodynamic links. While the signal power $\mu_b^2$ scales quadratically with control effort $I_b^2$ the noise variance grows quartically $I_b^4$ due to the coupling between emission and active propulsion. This leads to a SNR collapse at high intensities
\begin{equation}
    \text{SNR}(I) = \frac{\mu(I)^2}{\sigma_Y(I)^2} \approx \frac{(c_1 I)^2}{\sigma_m^2 + c_2 I^2 + c_3 I^4}.
    \label{eq:SNR_limit}
\end{equation}

\section{Detection and Bit Error Probability}
\label{sec:detection}

In this section we analytically derive the error performance of the chemo-hydrodynamic transceiver. We consider a binary modulation scheme wherein the control input $I_{\text{ext}}(t)$ is selected from the set $\{I_0, I_1\}$ with equiprobable occurrence over a symbol duration $T$.

\subsection{Hypothesis Testing Problem}
\label{subsec:hypotheses}

Predicated on the linearized observation model established in Section~\ref{sec:receiver_model} the received signal $Y$ follows a conditional Gaussian distribution
\begin{equation}
    \mathcal{H}_b: Y \sim \mathcal{N}(\mu_b, \sigma_b^2), \quad b \in \{0, 1\},
    \label{eq:hypotheses}
\end{equation}
where the mean $\mu_b$ and variance $\sigma_b^2$ depend on the control amplitude $I_b$ as follows
\begin{align}
\sigma_b^2 = \sigma_{Y,b}^2 \ \text{with}\ \sigma_{Y,b}^2 \text{ given by \eqref{eq:variance_intermediate}}.
\label{eq:sigma_b}
\end{align}
Combining \eqref{eq:variance_intermediate} with \eqref{eq:sigma_xb_exact} yields the polynomial form in \eqref{eq:variance_decomposition} (with coefficients depending on $T$ and $D_r$), which we use to obtain closed-form design rules.

Since the variance $\sigma_b^2$ is strictly increasing with $I_b$ the condition $\sigma_1 > \sigma_0$ holds for any signaling levels $I_1 > I_0 \ge 0$. This signal dependent noise constitutes a defining characteristic of the channel necessitating an optimal detection strategy that accounts for unequal variances.

\subsection{Maximum Likelihood Detection}
\label{subsec:ml_detector}

The ML detection strategy necessitates the evaluation of the likelihood ratio against unity which is mathematically equivalent to solving the quadratic equation
\begin{equation}
    \ln\left(\frac{\sigma_1}{\sigma_0}\right) + \frac{(Y-\mu_0)^2}{2\sigma_0^2} - \frac{(Y-\mu_1)^2}{2\sigma_1^2} = 0.
\end{equation}
While this equation generally yields two roots for the parameter regime of interest where $\mu_1 > \mu_0 \ge 0$ and $\sigma_1 > \sigma_0$ the lower root lies below $\mu_0$ in a region of negligible probability mass. Consequently the optimal decision rule simplifies to a single threshold test
\begin{equation}
    \hat{b} = \begin{cases} 1 & \text{if } Y > \gamma, \\ 0 & \text{if } Y \le \gamma. \end{cases}
    \label{eq:decision_rule}
\end{equation}
The optimal threshold $\gamma$ is given by the relevant root in the interval $[\mu_0, \mu_1]$
\begin{equation}
    \gamma = \frac{\mu_1 \sigma_0^2 - \mu_0 \sigma_1^2 + \sigma_0 \sigma_1 \sqrt{(\mu_1-\mu_0)^2 + 2(\sigma_1^2-\sigma_0^2)\ln(\sigma_1/\sigma_0)}}{\sigma_0^2 - \sigma_1^2}.
\end{equation}

\subsection{Bit Error Probability}
\label{subsec:bep}

The corresponding average Bit Error Probability BEP for this detection scheme is expressed as
\begin{equation}
    P_e = \frac{1}{2} Q\left( \frac{\gamma - \mu_0}{\sigma_0} \right) + \frac{1}{2} Q\left( \frac{\mu_1 - \gamma}{\sigma_1} \right),
    \label{eq:Pe_exact}
\end{equation}
where $Q(\cdot)$ denotes the standard Gaussian tail function.

\subsection{The Chemo Hydrodynamic trade-off}
\label{subsec:tradeoff}

In order to elucidate the fundamental channel constraints we analyze On Off Keying OOK where $I_0=0$ and $I_1=I$. We define a proxy SNR metric $\text{SNR}(I) = \mu_1^2/\sigma_1^2$ to expose the scaling behavior noting that the exact BEP is computed using the threshold rule in \eqref{eq:Pe_exact}
\begin{equation}
    \text{SNR}(I) = \frac{(c_1 I)^2}{\sigma_m^2 + c_2 I^2 + c_3 I^4},
    \label{eq:SNR_analysis}
\end{equation}
where $c_1$ $c_2$ and $c_3$ are physical constants derived from Eqs. \eqref{eq:mu_b} and \eqref{eq:sigma_b}.

Equation \eqref{eq:SNR_analysis} captures the joint sensing communication trade-off. At low amplitudes the SNR improves as $I^2$ because signal power dominates receiver noise $\sigma_m^2$. However at high amplitudes the active noise term $c_3 I^4$ dominates causing the SNR to collapse asymptotically as $I^{-2}$. This non monotonic behavior confirms the existence of a unique optimal control intensity $I_{\text{opt}}$.

\subsection{Theoretical Analysis of Optimal Control}
\label{subsec:optimization}

We ascertain the optimal actuation magnitude by solving the maximization condition $\frac{d}{dI} \text{SNR}(I) = 0$
\begin{equation}
    \frac{d}{dI} \text{SNR}(I) = \frac{2 c_1^2 I (\sigma_m^2 - c_3 I^4)}{(\sigma_m^2 + c_2 I^2 + c_3 I^4)^2}.
\end{equation}
Setting the numerator to zero yields the optimality condition
\begin{equation}
    \sigma_m^2 - c_3 I^4 = 0 \quad \Rightarrow \quad I_{\text{opt}} = \sqrt[4]{\frac{\sigma_m^2}{c_3}}.
    \label{eq:optimal_control}
\end{equation}
Substituting the physical parameters provides the closed form design rule
\begin{equation}
    I_{\text{opt}} = \left( \frac{\sigma_m^2 d^4}{H_0^2 K_{\text{control}}^2\, g(T)} \right)^{1/4}.
    \label{eq:I_opt_physical}
\end{equation}
Notably the optimal intensity depends only on the receiver noise floor and the quartic active noise coefficient while the quadratic passive diffusion term $c_2 I^2$ scales identically to the signal power and thus does not shift the stationary point affecting only the peak SNR value.

This result provides a concrete guideline for IoBNT system design given a reliability constraint the macro scale controller must restrict actuation to an envelope $I \le I_{\text{opt}}$ as exceeding this threshold yields diminishing returns and eventual link failure. Furthermore the scaling $I_{\text{opt}} \propto d$ implies that distant nodes can tolerate more aggressive propulsion whereas short range links require strictly limited control to mitigate motion induced fading.

\section{Numerical Analysis and Discussion}
\label{sec:results}

In this section, the analytical chemo-hydrodynamic framework is validated against stochastic Particle-Based Simulations (PBS), and the system performance is evaluated under realistic physical constraints.

\subsection{Simulation Framework and Parameters}
\label{subsec:simulation_setup}

To establish a rigorous ground truth for validation, we implemented a stochastic PBS that numerically integrates the full active Brownian dynamics, thereby bypassing the effective-diffusion (Gaussian displacement) and far-field linearization approximations utilized in the theoretical derivation. The particle trajectory is evolved via the Euler-Maruyama discretization of the Langevin equations derived in Section~\ref{sec:stochastic_propagation}:
\begin{align}
    \varphi_{k+1} &= \varphi_k + \sqrt{2 D_r \Delta t}\, \mathcal{N}(0,1), \\
    x_{k+1} &= x_k + U_b \cos(\varphi_k) \Delta t + \sqrt{2 D_t \Delta t}\, \mathcal{N}(0,1),
\end{align}
where $\Delta t$ denotes the simulation time step. We set $\Delta t = 10^{-4}\,\text{s}$ to ensure sufficient temporal resolution relative to the rotational relaxation time scale.

The physical parameters of the simulation environment are selected to rigorously replicate standard experimental setups for self-diffusiophoretic colloids \cite{Howse2007Selfmotile, DeGraaf2015diffusiophoretic}. Specifically, we model a polystyrene particle of radius $a=1\,\mu\mathrm{m}$ suspended in an aqueous medium at room temperature ($T_{\mathrm{env}}=293\,\mathrm{K}$, dynamic viscosity $\eta=10^{-3}\,\mathrm{Pa}\cdot\mathrm{s}$), yielding a characteristic rotational relaxation time of $\tau_r \approx 6\,\mathrm{s}$. The diffusion coefficients for both the solute and product species are fixed at $D_B=D_{sol}=2\times 10^{-9}\,\mathrm{m}^2/\mathrm{s}$, a value commensurate with small fuel molecules such as hydrogen peroxide \cite{Howse2007Selfmotile}. The effective propulsion mobility and base reaction rate are calibrated to yield propulsion velocities in the range of $1\text{--}5\,\mu\mathrm{m/s}$, reproducing the velocity-concentration scaling laws observed in the literature \cite{Zottl2023Modeling}.

For the communication link topology, the transmission distance is fixed at $d=50\,\mu\mathrm{m}$ with a symbol duration of $T=1\,\text{s}$, characteristic of short-range molecular communication scenarios \cite{mahfuz2015comprehensive}. The receiver noise is modeled as Additive White Gaussian Noise (AWGN) following the thermal noise framework \cite{kilinc2013receiver}. Consistent with thermal noise models for molecular receivers, we model the baseline electronic and counting noise $\sigma_m$ to yield a reference SNR of 20 dB in the absence of propulsion. This ensures that any reduction in link reliability observed in the highcontrol regime is strictly attributable to the stochastic active mobility of the transceiver.

\subsection{Validation of the Effective Channel Model}
\label{subsec:validation}
\begin{figure}[t]
    \centering
    \includegraphics[width = 0.5\textwidth]{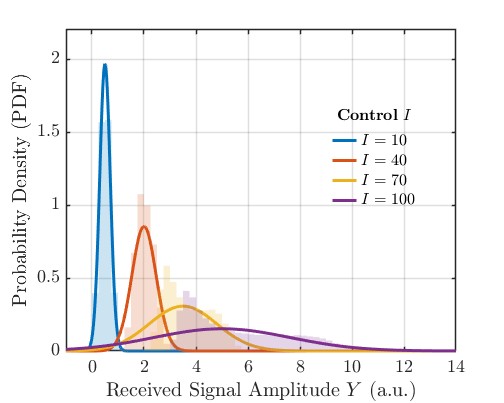}
    \caption{Model verification: analytical PDFs (solid) versus simulation histograms for $I\in\{10,40,70,100\}$. The widening distribution with increasing $I$ indicates dominant active-motion noise in water.}
\label{fig:model_verification}
\end{figure}
Unless stated otherwise, the simulation parameters satisfy the modeling assumptions in Sec.~II-A; here we validate the derived channel law against particle-based simulations.


Fig.~\ref{fig:model_verification} provides a comparative analysis between the analytical Gaussian probability density functions (PDFs) derived in Section~\ref{sec:detection} (solid lines) and the empirical histograms generated via the PBS (shaded bars) for four distinct control intensities $I \in \{10, 40, 70, 100\}$. Excellent agreement is observed across the entire spectrum of control inputs, thereby validating the accuracy of the proposed channel law.

The results elucidate two fundamental physical phenomena:
\begin{itemize}
    \item \textbf{Mean Shift:} As the control intensity $I$ increases, the centroid of the distribution shifts towards higher values, a direct consequence of the enhanced molecular emission rate ($\mu_Y \propto I$).
    \item \textbf{Variance Explosion:} Simultaneously, the distribution exhibits significant broadening. In the low-control regime ($I=10$, blue curve), the PDF remains narrow and peaked, indicating that the variance is dominated by the constant measurement noise floor $\sigma_m^2$. In stark contrast, within the high-control regime ($I=100$, purple curve), the distribution becomes markedly flat and dispersed. This behavior empirically corroborates the theoretical prediction that motion-induced variance scales quartically ($I^4$) with the control input.
\end{itemize}


\subsection{The Joint Trade-off}
\label{subsec:snr_tradeoff}

\begin{figure}[t]
    \centering
    \includegraphics[width = 0.48\textwidth]{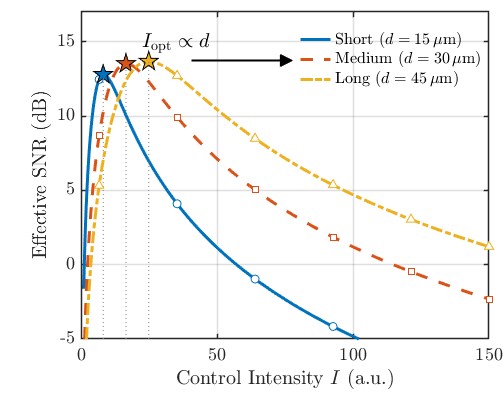}
    \caption{The joint sensing-communication trade-off. The effective SNR is plotted as a function of the control intensity $I$ for varying link distances $d$.}
\label{fig:snr_tradeoff}
\end{figure}

Figure~\ref{fig:snr_tradeoff} depicts the effective SNR as a function of the control intensity $I$ for a set of discrete link distances $d \in \{15, 30, 45\}\,\mu\text{m}$. The simulation results show a unimodal reliability profile governed by the coupling between mobility and transmission. In the low-control regime ($I < I_{\text{opt}}$), the SNR exhibits a quadratic improvement ($\text{SNR} \propto I^2$), as the enhanced molecular emission rate dominates the measurement noise floor. Conversely, beyond the optimal intensity threshold, the performance follows an asymptotic decay law of $\text{SNR} \propto I^{-2}$. This degradation confirms that excessive propulsion induces a quartic increase in motion-induced variance ($c_3 I^4$), which outweighs the benefits of signal amplification.


The analysis also indicates that the optimal control strategy is geometrically dependent. As marked by the star symbols in Fig.~\ref{fig:snr_tradeoff}, the optimal control intensity exhibits a linear scaling with the link distance ($I_{\text{opt}} \propto d$). This arises because the sensitivity gradient of the observation model scales as $\partial Y / \partial x \propto 1/d^2$. In the far-field regime, the received concentration profile becomes less sensitive to stochastic positional fluctuations, permitting a more aggressive propulsion strategy. Consequently, the peak SNR remains stable across the evaluated distances (varying by $<1\,\text{dB}$), suggesting that the system compensates for path loss through reduced motion sensitivity.

\subsection{Sensitivity and Robustness Analysis}
\label{subsec:sensitivity}

\begin{figure*}[t]
    \centering
    \includegraphics[width = 1\textwidth]{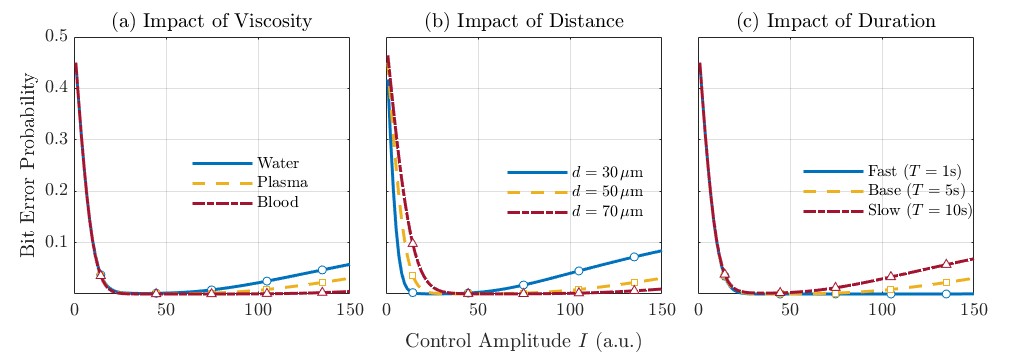}
\caption{Sensitivity analysis of the chemo-hydrodynamic link reliability. Bit Error Probability (BEP) versus control intensity $I$ for varying (a) fluid viscosity $\eta$, (b) link distance $d$, and (c) symbol duration $T$. The results quantify the performance degradation imposed by physiological environments (e.g., whole blood), geometric path loss, and accumulated diffusive jitter at lower data rates.}
\label{fig:sensitivity_analysis}
\end{figure*}

In order to rigorously evaluate the operational constraints of the transceiver architecture, Fig.~\ref{fig:sensitivity_analysis} delineates a comprehensive sensitivity analysis of the BEP with respect to fluid viscosity, link geometry, and symbol duration.

\subsubsection{Physiological Viscosity}
Figure~\ref{fig:sensitivity_analysis}(a) illustrates the system performance across a spectrum of biological fluids, ranging from cerebrospinal fluid ($\eta \approx 0.9\,\text{mPa}\cdot\text{s}$) to whole blood ($\eta \approx 3.5\,\text{mPa}\cdot\text{s}$). A discernible increase in the minimum achievable BEP is observed as viscosity increases. This phenomenon is governed by the interplay between two competing physical mechanisms: while elevated viscosity suppresses translational diffusion ($D_t \propto \eta^{-1}$), thereby theoretically stabilizing the positional variance, it concurrently suppresses rotational diffusion ($D_r \propto \eta^{-1}$). In the high-control regime, the latter effect becomes dominant, as a reduced $D_r$ extends the persistence length of the active motion ($U^2/D_r$), resulting in exacerbated directional deviations. Consequently, for reliable operation within whole blood environments, the control intensity must be attenuated relative to the aqueous optimum to mitigate this enhanced active noise contribution.

\subsubsection{Geometric Path Loss}
Figure~\ref{fig:sensitivity_analysis}(b) corroborates the fundamental range limitations of the system. As the link distance is extended from $30\,\mu\text{m}$ to $70\,\mu\text{m}$, the minimum BEP increases, and the optimal operating point exhibits a slight shift towards higher control magnitudes. This performance degradation is primarily attributable to the decay in signal amplitude ($\mu_Y \propto 1/d$), which precipitates an approximate $1/d^2$ scaling of the SNR. At extended ranges (represented by the red curve), the operational window for maintaining a low BEP becomes markedly constrained. This observation implies that a simple fixed-gain control strategy is inadequate for variable-range links; rather, distance-adaptive actuation is necessitated to ensure reliability at the network periphery.

\subsubsection{Diffusive Jitter Accumulation}
Figure~\ref{fig:sensitivity_analysis}(c) elucidates a temporal trade-off that is distinct from conventional static communication channels, where temporal averaging typically enhances the SNR. In the present chemo-hydrodynamic context, because the receiver samples the instantaneous concentration at time $T$, extending the symbol duration effectively degrades performance within the high-control regime. The "Fast" system configuration ($T=1\,\text{s}$) demonstrates superior performance compared to the "Slow" system ($T=10\,\text{s}$), a result stemming from the linear accumulation of position variance over time ($\sigma_x^2 \propto T$). Therefore, provided that the symbol duration satisfies the quasi-steady constraint ($(T \gtrsim d^2/D_B)$) essential for the validity of the channel model, higher data transmission rates are preferable to minimize the deleterious accumulation of motion-induced channel fading.

\subsection{Comparison with Standard Models}

\begin{figure*}[t]
    \centering
    \includegraphics[width = 0.85\textwidth]{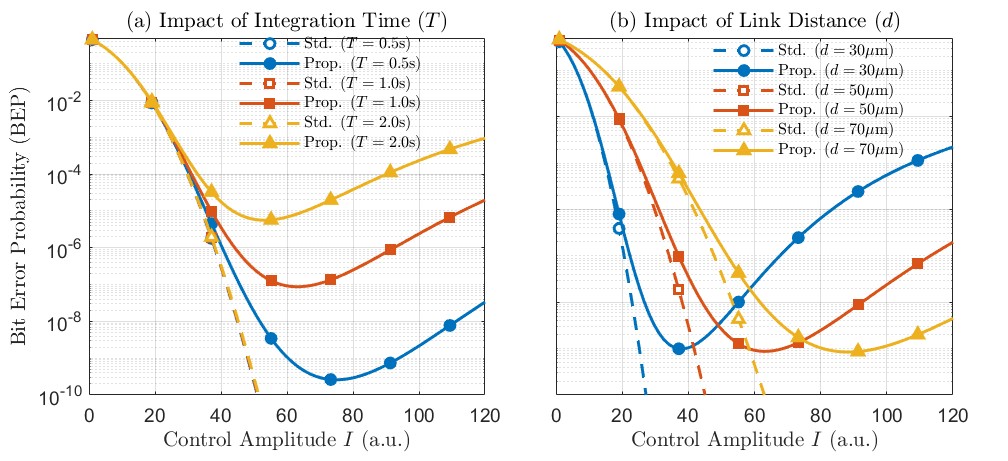}
    \caption{Impact of active mobility on reliability: BEP versus control intensity $I$ for the proposed chemo-hydrodynamic model (solid) and a Brownian baseline (dashed) with $D_{\text{eff}}\!\approx\! D_t$, adapted from \cite{Ahmadzadeh2018Stochastic}. (a) $T\in\{0.5,1.0,2.0\}\,\mathrm{s}$ at $d=50\,\mu\mathrm{m}$. (b) $d\in\{30,50,70\}\,\mu\mathrm{m}$ at $T=1.0\,\mathrm{s}$. Standard models miss the high-$I$ error rise due to propulsion-induced jitter.}

\label{fig:gap}
\end{figure*}

In order to rigorously quantify the estimation errors inherent in neglecting active motility noise, Fig.~\ref{fig:gap} presents a comparative analysis between the proposed chemo-hydrodynamic model (solid lines) and a standard diffusive baseline (dashed lines). Critically, the detection logic remains invariant across both scenarios, utilizing the same ML decision rule; the distinction lies exclusively in the mobility formulation. The baseline embodies the conventional modeling assumption wherein the transceiver's motion is approximated as independent Brownian motion ($D_{\text{eff}} \approx D_t$), thereby decoupling the diffusion dynamics from the control intensity.

\subsubsection{The Estimation Gap}
Figure~\ref{fig:gap} elucidates a fundamental divergence in the reliability predictions of the two paradigms. The standard model suggests that the BEP decreases monotonically with increasing control intensity $I$, eventually reaching saturation as the SNR becomes limited solely by passive diffusion. In stark contrast, the proposed chemo-hydrodynamic model reveals a non-monotonic reliability profile: beyond a critical intensity threshold, the BEP begins to deteriorate due to the quartic scaling of the active noise component ($U^2/D_r$). The vertical disparity between these curves constitutes an ``Estimation Gap,'' illustrating that standard diffusive abstractions can severely underestimate the error rate in high-mobility regimes, despite maintaining accuracy at lower propulsion levels.

\subsubsection{The Penalty of Integration Time}
Figure~\ref{fig:gap}(a) investigates the deleterious impact of extending the symbol duration $T$. While diffusive broadening inherently degrades the performance of both models as $T$ increases, the proposed model incurs a significantly more severe penalty. Specifically, the reliability curve for $T=2.0\,\text{s}$ (yellow solid line) exhibits a markedly higher minimum achievable BEP compared to the $T=0.5\,\text{s}$ case (blue solid line). This degradation is attributable to the temporal accumulation of active position variance, which scales as $\sigma_{\text{active}}^2 \approx U^2 T / D_r$. Consequently, a prolonged symbol duration permits the actively propelled transmitter to deviate more significantly from its expected trajectory, thereby exacerbating the discrepancy between the predicted and actual system performance.

\subsubsection{Vulnerability of Short Links}
Figure~\ref{fig:gap}(b) demonstrates that short-range links exhibit a paradoxical susceptibility to modeling inaccuracies. At a link distance of $d=30\,\mu\text{m}$ (blue curves), the standard model predicts a negligible error rate ($<10^{-6}$) in the high-intensity regime. However, the proposed model indicates that the error floor persists on the order of $10^{-3}$. This heightened sensitivity arises because the gradient of the received concentration with respect to position scales as $\partial Y / \partial x \propto 1/d^2$. In short-range scenarios, even minor propulsion-induced jitter induces substantial signal fluctuations. It is noteworthy that at extended distances (e.g., $d=70\,\mu\text{m}$), the predictions of the two models nearly converge in the vicinity of the optimal control point. This convergence reinforces the conclusion that the standard abstraction remains valid in far-field, low-speed scenarios but fails to capture the reliability dynamics of short-range links employing aggressive propulsion.

\section{Conclusion}
\label{sec:conclusion}

In this paper, we developed a unified chemo-hydrodynamic physical-layer model for mobile IoBNT links, focusing on catalytic Janus transceivers where a single external control signal governs both propulsion and molecular emission. By capturing the intrinsic coupling between mobility and the received concentration statistics, we showed that the actuation mechanism simultaneously drives the mean signal and introduces a dominant mobility-induced noise term: the received mean increases linearly with control intensity, while the variance contains an active component that scales quartically with the control input.

This input-dependent noise yields a non-monotonic reliability profile, implying that overly aggressive actuation can push the link into a high-error regime. We derived closed-form design rules for the optimal actuation level as a function of link distance, and clarified how extended symbol durations under instantaneous sampling can be limited by accumulated motility-induced jitter. Particle-based simulations corroborated the proposed channel law and quantified a significant ``Estimation Gap,'' showing that standard decoupled Brownian-mobility models can severely underestimate the error probability in high-mobility regimes.

These findings provide physical-layer guidelines for mobility-aware IoBNT design, including actuation selection, symbol-duration choice, and reliability-targeted link adaptation. Future work will extend the framework to 3D geometries with flow, multi-node coordination, and experimental validation with fabricated Janus micromotors.

\appendices
\section{Chemo-Hydrodynamic Derivation}
\label{app:chemo_derivation}
This appendix derives the propulsion law used in Sec.~II by projecting the controlled surface flux onto the first Legendre mode. We use $\mu\triangleq\cos\theta$ to avoid confusion with the axial position $x(t)$.

\subsection{Legendre Projection}

We characterize the catalytic cap as imposing an inward (into the surface) consumption flux of fuel $A$ on the particle boundary. Consistent with Sec.~II, we write the boundary condition as $D\,\partial_r c_A\vert_{r=a}=k(\theta,t)$, where $k(\theta,t)\ge 0$ denotes the prescribed inward consumption flux on the catalytic domain. The cap indicator function satisfies
\begin{equation}
\Theta(\alpha-\theta)=\Theta(\cos\theta-\cos\alpha)=\Theta(\mu-\cos\alpha).
\end{equation}

The axisymmetric concentration field admits a series expansion in terms of Legendre polynomials as
\begin{equation}
c(r,\theta,t)=\sum_{\ell=0}^\infty A_\ell(t)\left(\frac{a}{r}\right)^{\ell+1}P_\ell(\mu).
\end{equation}
Since only the $\ell=1$ mode contributes to net translation for a force-free sphere, we restrict the projection to $A_1(t)$. Substituting the expansion into the flux boundary condition and projecting onto $P_1(\mu)$ yields the coefficient $A_1(t)$.

Invoking the orthogonality of Legendre polynomials introduces the normalization factor $3/2$ for $\ell=1$. This procedure yields
\begin{equation}
A_1(t)
=-\frac{a\kappa_{\text{base}}I_{\text{ext}}(t)}{D}\,\frac{3}{4}
\int_{\cos\alpha}^{1}P_1(\mu)\,d\mu.
\label{eq:A1_proj_app}
\end{equation}
Given the identity $P_1(\mu)=\mu$ the integral evaluates to $\int_{\cos\alpha}^{1}\mu\,d\mu=\frac{1}{2}\sin^{2}\alpha$. Consequently we obtain
\begin{equation}
A_1(t)=-\frac{3a}{8D}\sin^{2}\alpha\,\kappa_{\text{base}}I_{\text{ext}}(t).
\label{eq:A1_app}
\end{equation}

\subsection{Slip Mode \texorpdfstring{$B_1$}{B1}}

The tangential phoretic slip velocity is defined as $\mathbf{v}_s=-b\nabla_{\parallel}c\vert_{r=a}$ or equivalently expressed as
\begin{equation}
\mathbf{v}_s(\theta,t)=-\frac{b}{a}\,\frac{\partial c}{\partial\theta}\,\hat{\mathbf{e}}_{\theta}.
\end{equation}
Restricting the analysis to the primary mode contribution yields
\begin{equation}
\frac{\partial c}{\partial\theta}\approx A_1(t)\frac{\partial P_1(\mu)}{\partial\theta}
=A_1(t)\left(-\sin\theta\right),
\label{eq:dc_dtheta_app}
\end{equation}
such that the tangential component becomes $v_{s,\theta}\approx \frac{b}{a}A_1(t)\sin\theta$. By identifying the squirmer coefficient as $B_1(t)=\frac{b}{a}A_1(t)$ and substituting \eqref{eq:A1_app} we recover
\begin{equation}
B_1(t)=-\frac{3b_{\mathrm{DP}}}{8D}\sin^{2}\alpha\,\kappa_{\text{base}}I_{\text{ext}}(t),
\end{equation}
a result that corresponds directly to \eqref{eq:B1_main} presented in the main text.

\bibliographystyle{IEEEtran}
\bibliography{references}

\vfill
\end{document}